\documentclass[%
 reprint,
superscriptaddress,
 amsmath,amssymb,
 aps,
 prl,
 showkeys,
]{revtex4-2}

\usepackage{graphicx}
\usepackage{dcolumn}
\usepackage{bm}
\usepackage{siunitx}
\usepackage{physics}
\usepackage[dvipsnames]{xcolor}
\usepackage{tikz} 
\usepackage{float}
\usepackage[breaklinks=true, hidelinks]{hyperref}

\definecolor{drkgreen}{rgb}{0.0, 0.5, 0.0}
\definecolor{violet}{rgb}{0.5, 0.0, 1.0}

\newcommand{\bl}[1]{\textcolor{blue}{#1}}

\begin{document}

\preprint{APS/123-QED}

\title{Continuous-variable square-ladder cluster states in a microwave frequency comb}

\author{Fabio Lingua}
 \email{lingua@kth.se}
 \affiliation{Department of Applied Physics, KTH Royal Institute of Technology, SE-10691 Stockholm, Sweden}
 
\author{J. C. Rivera Hernández}
 \affiliation{Department of Applied Physics, KTH Royal Institute of Technology, SE-10691 Stockholm, Sweden}

\author{Michele Cortinovis}
 \affiliation{Department of Applied Physics, KTH Royal Institute of Technology, SE-10691 Stockholm, Sweden}
 \affiliation{Dipartimento di Fisica, Politecnico di Milano, I-20133 Milano, Italy}

\author{David B. Haviland}
 \affiliation{Department of Applied Physics, KTH Royal Institute of Technology, SE-10691 Stockholm, Sweden}

\date{\today}

\begin{abstract}
We describe an experiment demonstrating the generation of three independent square-ladder continuous-variable cluster states with up to 94 qumodes of a microwave frequency comb. 
This entanglement structure at a large scale is realized by injecting vacuum fluctuations into a Josephson Parametric Amplifier pumped by three coherent signals around twice its resonance frequency, each having a particular well-defined phase relation.
We reach up to \SI{1.4}{dB} of squeezing of the nullifier which verifies the cluster state on the square ladder graph. 
Our results are consistent with a more familiar measure of two-mode squeezing, where we find up to \SI{5.42}{dB} for one pump, and up to \SI{1}{dB} for three pumps.
\end{abstract}

\maketitle



Measurement-based quantum computation (MBQC) is a universal scheme~\cite{raussendorf_one-way_2001} where algorithms are executed through sequential measurements on a highly entangled resource state, known as a cluster state~\cite{briegel_measurement-based_2009}.
While MBQC was originally formulated for discrete variables (qubits), continuous-variable (CV) quantum information processing with their inherent high dimensional Hilbert spaces and efficient scaling to many degrees of freedom, has emerged as a strong alternative~\cite{braunstein_quantum_2005, menicucci_graphical_2011, pfister_continuous-variable_2020}.
CV quantum information is encoded in the quadrature eigenstates, called \emph{qumodes}.
CV cluster states are multipartite entangled states defined as the continuous superposition of qumodes, where entanglement is established through correlations between their quadratures, typically verified through squeezing measurements.
Realizing a well-controlled structure of quantum correlations across a large number of qumodes is imperative for the practical implementation of CV MBQC.

Cluster state entanglement is characterized by a well-defined topology of correlations, depicted as a lattice graph with nearest neighbor links between qumodes~\cite{pfister_continuous-variable_2020}.
Verification of the multipartite entanglement necessary for effective computation is also a major challenge.
When the number of qumodes in the system increases, verifying entanglement through traditional bipartition tests~\cite{simon_peres-horodecki_2000, shchukin_generalized_2015} becomes impossible, as the number of bipartitions scales exponentially with the size of the system.
Variance-based entanglement tests that target specific correlation topologies are multi-mode generalizations of two-mode squeezing (EPR entanglement~\cite{ou_realization_1992}).
Such tests are indispensable to the verification of entanglement for large CV systems~\cite{hyllus_optimal_2006, menicucci_graphical_2011, pfister_continuous-variable_2020}.

CV cluster states have been successfully realized with optical frequency combs on a small to intermediate scale~\cite{pysher_parallel_2011, cai_multimode_2017, walschaers_emergent_2023, roman-rodriguez_multimode_2024}. 
A design for arbitrarily large CV cluster states has been proposed~\cite{menicucci_arbitrarily_2010}, and physically realized with time-multiplexing techniques~\cite{menicucci_temporal-mode_2011, chen_experimental_2014, asavanant_time-domain-multiplexed_2021, du_generation_2023} enabling quantum advantage~\cite{madsen_quantum_2022}.
\begin{figure}[H]
    \includegraphics[width=\columnwidth]{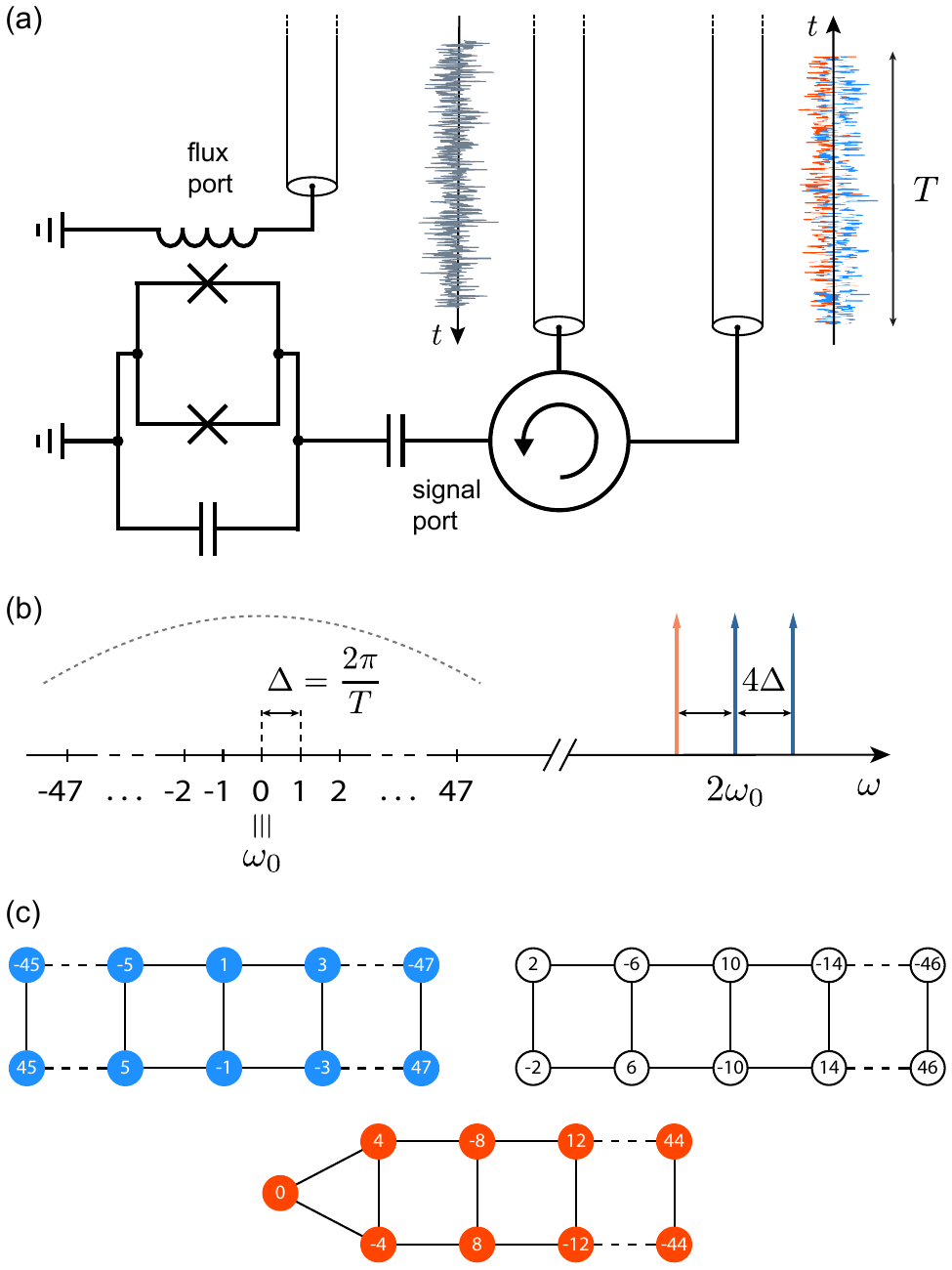}
    \caption{\label{fig:JPA} (a) Schematic of the experiment.
    (b) The measurement basis and the pumping scheme in the frequency domain. 
    The output signal is sampled at $95$ qumodes equally spaced by $\Delta=2\pi \times \SI{0.1}{\mega\hertz}$ placed and labeled symmetrically around $\omega_0$ (qumode $0$).
    Three coherent signals pump the JPA at $2\omega_0$ and $2\omega_0\pm 4\Delta$ $\left(\omega_0=2\pi \times\SI{4.2}{\giga\hertz}\right)$.
    The lowest-frequency pump (orange) is set with a phase $\pi$, while the other two pumps (blue) are set to phase $0$.
    (c) The three canonical graphs corresponding to the three measured cluster states comprising the 48 odd qumodes (blue), and the 24 (white) and 23 (red) even qumodes respectively.}
\end{figure}
\noindent
Digital signal processing at microwave frequencies and quantum circuits operating at millikelvin temperatures provide a new platform for CV quantum computation~\cite{jing_experimental_2006, hung_quantum_2021, petrovnin_generation_2023, jolin_multipartite_2023}.
Microwave CV cluster states have been realized between a limited number of qumodes~\cite{hung_quantum_2021, petrovnin_generation_2023}, but the realization at a large scale is still lacking.

Here we build on our previous work with digital control of multi-modal scattering in a microwave frequency comb~\cite{rivera_hernandez_control_2024} to demonstrate the entanglement of vacuum fluctuations by a Josephson Parametric Amplifier (JPA).
The JPA is pumped with three coherent tones having a particular phase configuration.
We realize three independent CV square-ladder cluster states spanning 94 qumodes of a frequency comb.


Our experiment measures the noise quadratures of vacuum fluctuations that scatter off of a JPA (see FIG.~\ref{fig:JPA}(a)).  
The JPA operates as a parametric oscillator, comprising a lumped-element LC circuit with a superconducting quantum interference device (SQUID) acting as a flux-tunable inductor.
Inductive coupling of the flux port to the SQUID loop allows a time-varying flux pump to modulate the circuit's inductance, facilitating frequency mixing (intermodulation) with signals entering through the signal port.
Without flux the JPA's resonance frequency is \SI{7.8}{\giga\hertz}.
Applying a DC flux bias of approximately $\pm 0.48\Phi_0$ shifts the resonance frequency $\omega_0 /2\pi$ to about \SI{4.2}{\giga\hertz}.
The JPA is cooled to \SI{10}{\milli\kelvin} and overcoupled at the signal port to a circulator, giving a loaded quality factor $Q = 37.5$, corresponding to a linewidth $\kappa = 2\pi \times \SI{112}{\mega\hertz}$.
The circulator separates the incoming vacuum fluctuations from the outgoing correlated vacuum, which is routed through a double isolator to a cryogenic low-noise amplifier.
A complete schematic diagram of the measurement setup is provided in the Supplemental Material~\cite{lingua_supplemental_2024}.

After amplification, the correlated noise is sent to a digital multifrequency lock-in amplifier for demodulation~\cite{tholen_measurement_2022}. 
The pump generation is also managed by the multi-frequency lock-in which is capable of simultaneous coherent modulation and demodulation at up to 192 frequencies.
The frequencies of the pump and demodulation tones are chosen from a basis comprising integer multiples of the measurement bandwidth $\Delta = 2\pi/T$, with $T$ being the measurement time window.
The integer-multiple condition defines an orthonormal basis set of qumodes $\{q_i\}$, and a global phase reference, i.e. the phase of a signal at frequency $\Delta$.  
The measurement bandwidth $\Delta$ is tuned to be commensurate with the digital sampling frequency, to eliminate Fourier leakage between demodulated qumodes.
In the experiment described here we select $\Delta = 2\pi \times \SI{0.1}{\mega\hertz} \ll  \kappa$, and perform quadrature demodulation across a basis of 95 tones spanning \bl{\SI{9.5}{\mega\hertz}}, all well within the bandwidth of the JPA.
This constitutes our frequency comb of qumodes.
We center and label the comb symmetrically around $\omega_0$ (qumode $0$), as depicted in FIG.~\ref{fig:JPA}(b).

We drive the JPA with three pumps of equal strength at frequencies $2\omega_0$ and $2\omega_0 \pm 4\Delta$, as depicted in FIG.~\ref{fig:JPA}(b).
The frequency of the central pump tone is restricted by the tuning of the mode basis, but placed as close as possible to twice the JPA resonant frequency $2\omega_0$.  
The pumps at $2\omega_0$ and $2\omega_0 + 4\Delta$ (blue arrows) have phase zero, while the pump at $2\omega_0-4\Delta$ (orange arrow) is set to $\pi$.
This pump configuration creates a square lattice structure of the mode correlations, as shown in our previous work~\cite{rivera_hernandez_control_2024}.

We measure the output voltage noise quadratures at each frequency in our mode basis which we denote $x_i, p_i$ (units Volts) and construct the measured covariance matrix $V^\text{meas}$ in units of photon number,
\begin{equation}
    V^\text{meas}_{ij} = \dfrac{ \expval{\Delta q_i \Delta q_j}}{Z_c \hbar \Delta \sqrt{\omega_i \omega_j}} = \dfrac{ \expval{\left( q_i - \expval{q_i} \right) \left( q_j - \expval{q_j} \right)}}{Z_c \hbar \Delta \sqrt{\omega_i \omega_j}}.
    \label{eq:cov-mat}
\end{equation}
Here $q_i=x_i, p_i$ are the measured quadratures of the $i$-th mode, $\expval{q_i}$ their expectation values, and $Z_c=\SI{50}{\ohm}$ is the impedance loading the signal port.
The enumeration scheme places the qumode $i=0$ at $\omega_0$.
To recover the covariance matrix of the output qumodes $V^\text{quant}$, we must remove the uncorrelated classical added noise~\cite{jolin_multipartite_2023}, which requires calibration of the amplification chain~\cite{mariantoni_planck_2010}.
After this removal we ensure that $V^\text{quant}$ is physical, i.e. satisfies the Heisenberg uncertainty principle, by performing a constrained minimization~\cite{shchukin_recovering_2016} obtaining the physical covariance matrix $V$.
Details on the calibration of the amplification chain and the reconstruction of physical covariance matrices are provided in the Supplemental Material~\cite{lingua_supplemental_2024}.

\begin{figure}
    \includegraphics[width=\columnwidth]{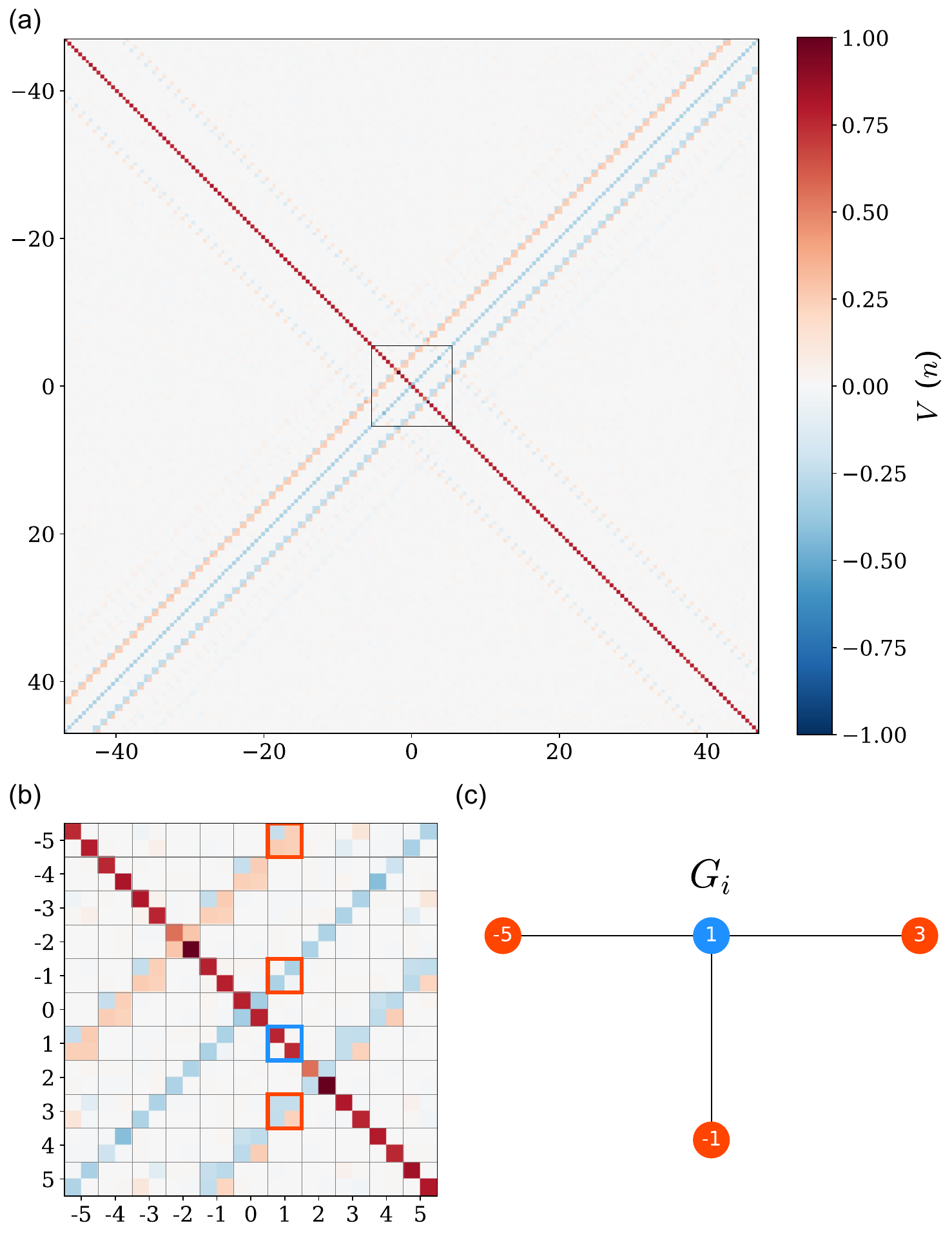}
    \caption{\label{fig:cov} (a) Covariance matrix $V$ at pump power $g=0.30g_\text{3dB}$. 
    (b) Zoom of $V$ between qumodes $-5$ and $5$. 
    Blue and red boxes highlight the non-zero $2 \cross 2$ blocks $V_{ij}$ used to generate a single-mode canonical graph.
    A qumode $i$ (blue) is connected to qumodes $j$ (red) featuring non-vanishing $V_{ij}$.
    (c) Single-mode canonical graph $G_i$ of qumode $i=1$.}
\end{figure}

Figure~\ref{fig:cov}(a) shows the covariance matrix $V$ in units of photon number for pump power $g=0.30g_\text{3dB}$, where $g_\text{3dB}$ represents the pump power for \SI{3}{\dB} gain of an injected signal.
The main diagonal elements give the quadrature variances $\Delta x_i^2$, $\Delta p_i^2$ and the three anti-diagonals give inter-mode covariances due to the 2nd order intermodulation processes induced by the pumps~\cite{rivera_hernandez_control_2024}.
Figure~\ref{fig:cov}(b) shows a zoom of the central region of the covariance matrix between qumodes $-5$ and $5$.
Non-zero off-diagonal $2 \cross 2$ blocks contain the quadrature covariances $\Delta q_i\Delta q_j$ containing information about squeezing between qumode pairs $i$ and $j$.
Positive elements (red) represent correlations, whereas negative elements (blue) represent anti-correlations.
Following similar arguments as in~\cite{jolin_multipartite_2023, rivera_hernandez_control_2024} we recover the single-mode canonical graph $G_i$ shown in FIG.~\ref{fig:cov}(c), representing the structure of correlations with mode $i$.
For example qumode $i=1$ in Fig.~\ref{fig:cov}(b) (blue block) is connected to qumodes $j=\{-5,-1, 3\}$ as witnessed by a non-vanishing covariance (red blocks).

The covariance matrix exhibits global symmetries: (i) qumode $i$ is always correlated to qumodes ${j=\{-i, -i\pm4\Delta\}}$ reflecting the signal-idler intermodulation processes; (ii) the covariance of each quadrature pair ($2 \cross 2$ block) is invariant under translation of $i$ (i.e., every mode $i$ experiences the same type of quadrature correlations with respect to each of its relative neighbors $j\in G_i$).
Both symmetries are preserved under a global phase rotation corresponding to a change of the measurement basis.
Such a global rotation is obtained by individually rotating all the mode quadratures by the same angle $\theta$ (see supplemental material~\cite{lingua_supplemental_2024} for details).

Assembling single-mode graphs $G_i$ for all qumodes $i$ we recover the three independent square-ladder correlation graphs in FIG.~\ref{fig:JPA}(c). 
They correspond to the canonical graph of three CV cluster states involving the 48 odd qumodes (blue), and the 24 (white) and 23 (red) even qumodes.
The combined effect of the pumps links every other qumode, effectively dividing the full set into two primary subsets: even-indexed and odd-indexed.
Furthermore, within the even subset, the presence of the degenerate qumode $0$ induces an additional splitting, resulting in three independent cluster states.

CV cluster states are represented by a canonical graph featuring a lattice topology~\cite{pfister_continuous-variable_2020}.
The state is formally defined as a continuous superposition of $n$ qumodes $\ket{x_i}$ on which a CV $C_z$ gate is applied between qumode pairs
\begin{equation}
    \ket{\Psi}=\frac{1}{2\pi}\iint d\vec{x} \prod_{i,j\in G} e^{i\alpha_{ij} x_i x_j} \ket{x_1}\ket{x_2}\dots\ket{x_n}.
    \label{eq:CS_def}
\end{equation}
Here the gate $C_{z,\, ij}=e^{i\alpha_{ij} \hat{x}_i \hat{x}_j}$ with $ij$ denoting vertices of the graph $G$, and $\alpha_{ij}\in\mathbb{R}$.

We perform a variance-based nullifier test to prove that the measured covariance of the multi-modal squeezed vacuum on the canonical graph represents a CV cluster state.
The symmetry of the cluster state defines the nullifier.
Given a cluster state $\ket{\Psi}$ there exists a transformation $S$ called stabilizer, such that the cluster state is invariant under $S$
\begin{equation}
    S\ket{\Psi}=\ket{\Psi}.
    \label{eq:CStab}
\end{equation}
The global stablizer $S = \bigoplus_{i=1}^n S_i$, where the local stabilizers $S_i=e^{i\alpha N_i}$ are defined as exponentiation of the nullifier operator
\begin{equation}
    N_i=\hat{p}_i - \sum_{j\in G_i}h_{ij} \hat{x}_j.
    \label{eq:Null}
\end{equation}
Here $\hat{p}_i$, $\hat{x}_j$ are quadrature operators, $h_{ij}\in\mathbb{R}$, and the sum on $j$ runs on the qumodes connected to $i$ in the single-mode graph $G_i$.
Choosing $h_{ij}=\alpha_{ij}$ ensures the stabilization condition~\eqref{eq:CStab} and uniquely defines the CV Pauli $X$ and $Z$ gates within the gate set of MBQC (see supplemental material~\cite{lingua_supplemental_2024} for details).

In the covariance matrix shown in FIG.~\ref{fig:cov}(a)-(b), the antidiagonal blocks $V_{i,-i+4}$ have opposite sign to $V_{i,-i-4}$ and $V_{i,-i}$.
This change of sign is intimately connected with the phase of the left (orange) pump in FIG.~\ref{fig:JPA}(b) being set to $\pi$, suggesting the choice $h_{ij}=\pm\left\{1,-1,-1\right\}$ for $j=\left\{-i+4,-i,-i-4\right\}$ respectively.

\begin{figure}
\includegraphics[width=\columnwidth]{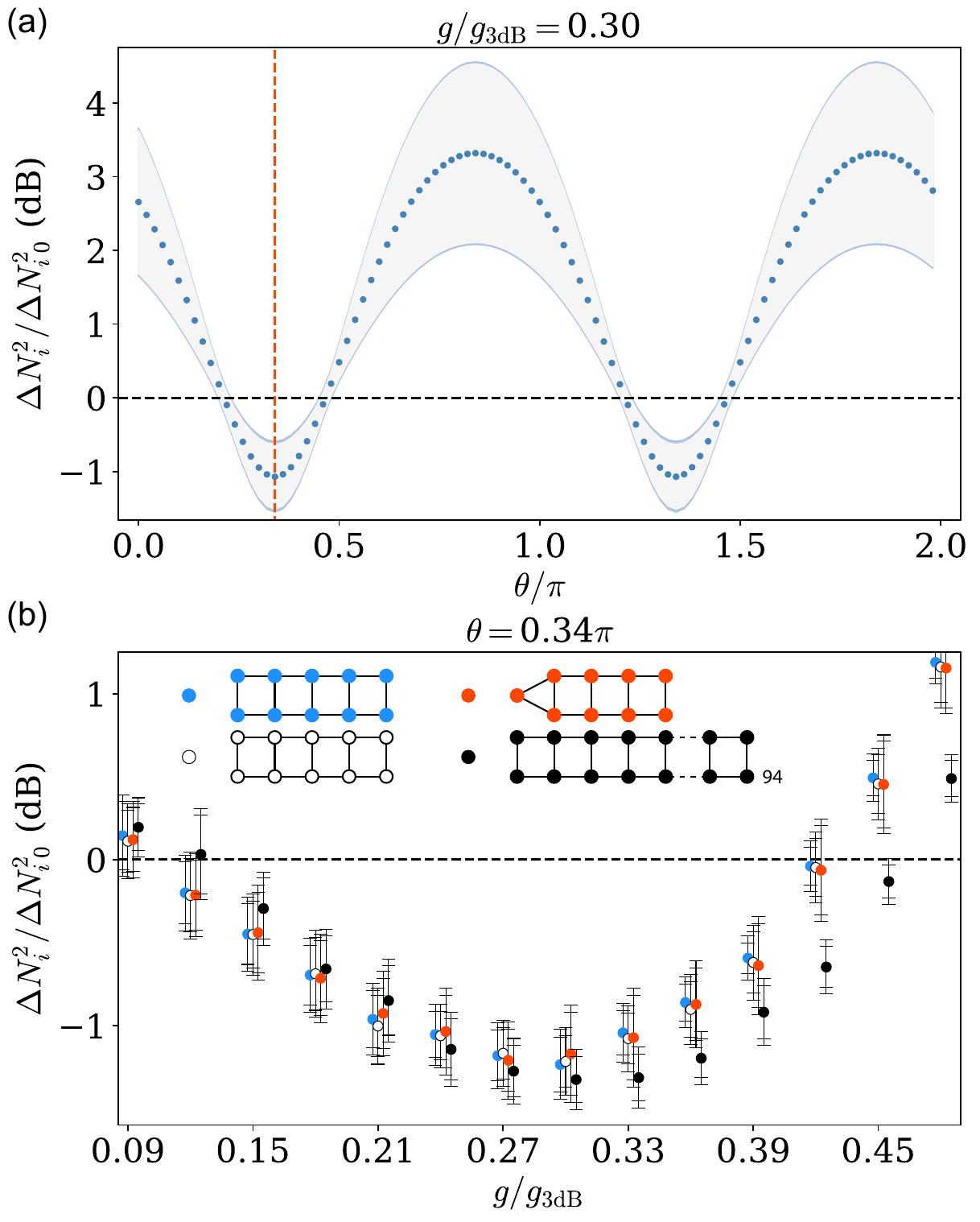}
    \caption{\label{fig:null} (a) Nullifier variance in \SI{}{\dB} as a function of the global rotation angle $\theta$ at $g=0.30g_\text{3dB}$.
    Maximum squeezing is achieved for $\theta \approx 0.34(k+1)\pi$, $k\in\mathbb{Z}$.
    The light gray area shows the standard deviation among qumodes, bounded by a dark gray area showing the measurement uncertainty.
    (b) Nullifier variance in \SI{}{\dB} for each subgraph as a function of pump power $g/g_\text{3dB}$ at the angle of maximum squeezing [$\theta$ corresponding to the vertical red dashed line in panel (a)].
    Black dots refer to the blue subgraph extended to include 94 odd modes.
    The error bars show the standard deviation between qumodes (inner error bars) plus the estimated measurement uncertainty (outer error bars)}.
\end{figure}

A cluster state $\ket{\Psi}$ described by the graph $G$ has an expectation value $\expval{N_i}=0$, and variances $\Delta N_i^2<\Delta N_{i\;0}^2$, where $\Delta N_{i\;0}^2\equiv\ev{\Delta N_{i}^2}{0}$ is the variance of the nullifier~\eqref{eq:Null} measured on vacuum $\ket{0}$, i.e. in the absence of pumping.
Figure~\ref{fig:null}(a) shows $\Delta N_i^2/\Delta N_{i\,0}^2$ as a function of the global phase rotation $\theta$.
Each point marks the average over the qumodes of all three canonical graphs.
The light-gray shaded area gives the standard deviation of the nullifier, bounded by a thin dark-gray area representing the estimated measurement uncertainty.
The nullifer exhibits up to \SI{1}{\dB} squeezing at $\theta \approx 0.34(k+1)\pi$, $k\in\mathbb{Z}$ (see supplemental material~\cite{lingua_supplemental_2024} for details).  
In FIG.~\ref{fig:null}(b) we show how the nullifier variance changes with pump power at the angle of maximum squeezing [the value of $\theta$ represented by the vertical red dashed line in Fig.~\ref{fig:null}(a)].
We see a global minimum of $\Delta N_i^2/\Delta N_{i\,0}^2$ for a pump power of $g\approx 0.30g_\text{3dB}$.
The inner error bars refer to the standard deviation between qumodes of the same graph, while the outer error bars include the estimated measurement uncertainty.

A nullifier below zero reflects the ability of our experimental setup to squeeze vacuum~\cite{menicucci_graphical_2011, pfister_continuous-variable_2020}, which may be limited by several factors including internal loss in the JPA and insertion loss of the circulator and cabling.
Furthermore, our nullifier does not account for higher-order mixing processes which are much weaker, but nevertheless induce correlations that are not in our canonical graph $G_i$~\cite{gonzalez-arciniegas_cluster_2021}.
Indeed at higher pumping power, we find that two off-diagonals corresponding to these higher-order mixing processes, at $i,\,i\pm 8$, rise above the measurement noise floor.
These additional mode couplings degrade the squeezing of the measured nullifier.

\begin{figure}
\includegraphics[width=\columnwidth]{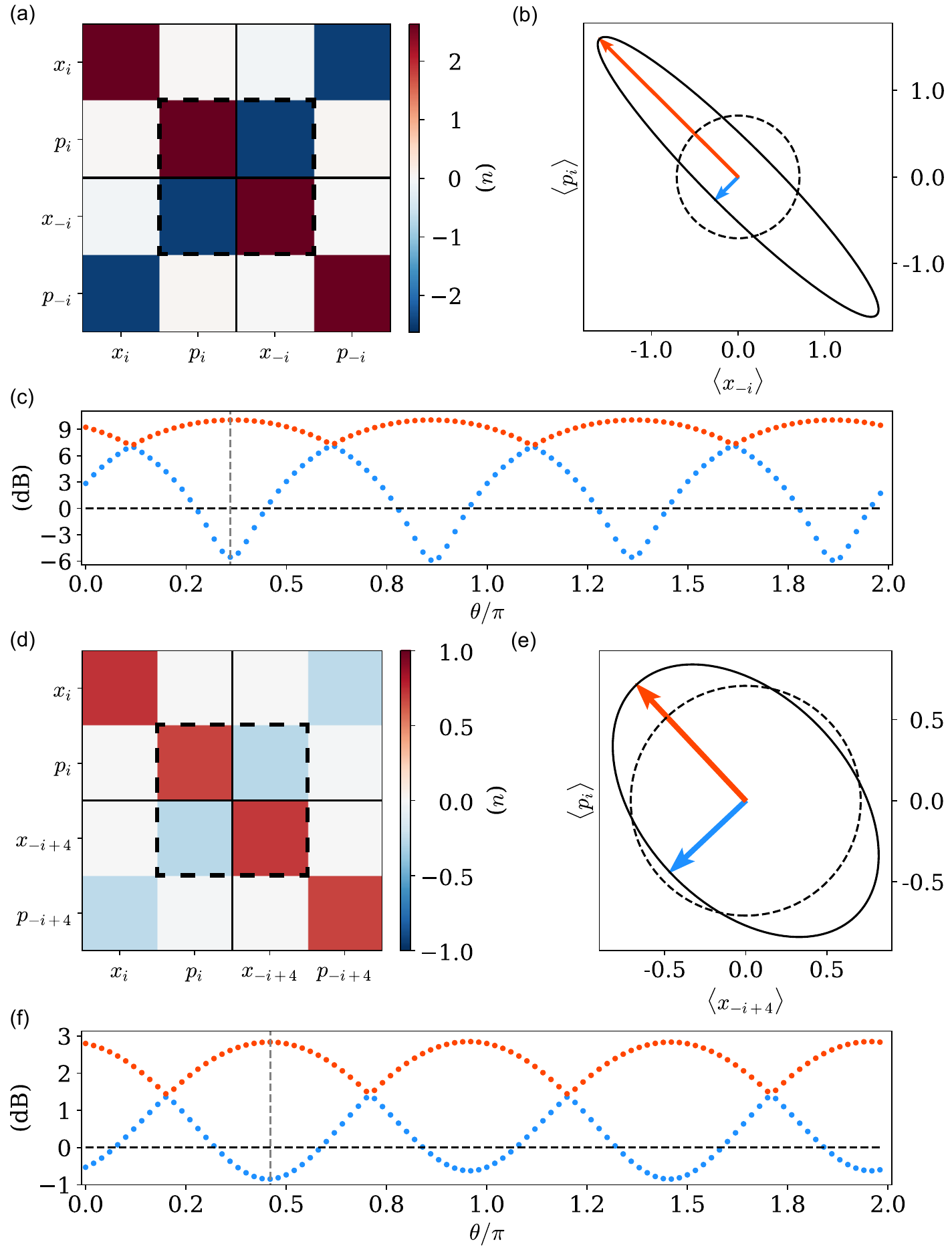}
    \caption{\label{fig:quad_sqz} Two-mode squeezing for the one pump (a)-(c) and three pumps (d)-(f).
    (a) and (d) show the two-mode covariance matrix between $i,j=-i$ and $i,j=-i + 4$ respectively.
    (b) and (e) compare the squeezing ellipses computed for the $\theta$ of maximum squeezing [vertical dashed gray line in (c) and (f)]. (c) and (f) show the eigenvalues of the $2\cross 2$ covariance [dashed box in (a)  and (d)] as a function of the global rotation angle $\theta$. 
    }
\end{figure}

To quantify squeezing in our experimental setup we compare the nullifier of our three-pump scheme with that of the simpler and more standard case of two-mode squeezing, where the JPA is pumped with a single tone at $2\omega_0$ [only the center pump in Fig.~\ref{fig:JPA}(b)].
With one pump [see FIG.~\ref{fig:quad_sqz}(a)-(c)] the covariance matrix has only one anti-diagonal of idlers, reflecting multiple two-mode squeezed states (see Supplemental material~\cite{lingua_supplemental_2024}).
The simplified correlation graph connects multiple mode pairs $i$ and $-i$ and the definition of the nullifier on this graph results in the expression $N_i=\hat{p}_i-\hat{x}_{-i}$, which is a familiar measure of two-mode squeezing.

In FIG.~\ref{fig:quad_sqz}(a) we show the two-mode covariance matrix between correlated qumodes $i$ and $-i$ in the case of one parametric pump for the rotation indicated by the vertical dashed line of FIG.~\ref{fig:quad_sqz}(c).
We compute the eigenvalues and eigenvectors of the relevant $2 \cross 2$ sub-covariance matrix (dashed box) containing the quadratures $\hat{p}_i$ and $\hat{x}_{-i}$. 
Figure~\ref{fig:quad_sqz}(c) shows these eigenvalues as a function of the global rotation angle $\theta$.
The vertical gray dashed line indicates the angle of maximum squeezing, represented by the eigenvectors and ellipse shown in FIG.~\ref{fig:quad_sqz}(b).
We reach on average \SI{-5}{\dB} of two-mode squeezing, comparable to~\cite{castellanos-beltran_amplification_2008, mallet_quantum_2011, zhong_squeezing_2013, malnou_optimal_2018}.

Figure~\ref{fig:quad_sqz}(d)-(f) show the same two-mode squeezing measurements between qumodes $i$ and $-i+4$ for the three-pump case.
While it is possible to reach up to \SI{-5.42}{\dB} of two-mode squeezing with a single pump, this level is reduced to maximum \SI{-1}{\dB} with three pumps.
The observed reduction confirms that additional mode couplings degrade two-mode squeezing.
We note that with either one or three pumps, the minimum variance of the nullifier in FIG.~\ref{fig:null} is comparable to the maximum two-mode squeezing in FIG.~\ref{fig:quad_sqz}. 
Furthermore, the imbalance between squeezing and anti-squeezing seen in FIG.~\ref{fig:quad_sqz}(c) and (f) confirms the presence of losses in our setup.
The losses also explain the degradation of squeezing with increasing pump power (see numerical analysis in the supplemental material~\cite{lingua_supplemental_2024}).

Finally, we reprogrammed the demodulation tones of our multi-frequency lock-in amplifier to measure the quadrature noise at 94 frequencies corresponding to the odds qumodes on the blue subgraph in FIG.\ref{fig:JPA}(c). 
We set our $94$ demodulators to measure only at the frequencies of odd qumodes, not measuring the even qumodes.
Thus we extend the blue canonical graph from $[-47,47]$ to $[-95,95]$.
With the same three-pump scheme we generate a single cluster state and perform the nullifier test on all 94 modes.
We find a slightly improved squeezing of the nullifer of \SI{-1.4}{dB} [see black points in FIG.~\ref{fig:null}(b)].


In summary, we demonstrate the power of microwave digital signal processing to generate and measure large-scale entanglement in the frequency domain.
Using three pumps with a specific phase configuration to interfere idlers in a JPA, we create quantum correlations described by a square-ladder graph.
Through analysis of the covariance matrix when the vacuum is input to the JPA, we confirmed the generation of three independent square-ladder continuous-variable cluster states.
While the nullifier defined on the cluster state of 94 qumodes shows \SI{-1.4}{dB} of squeezing, the two-mode cluster states show \SI{-5.42}{dB} of squeezing.
Through further engineering of the losses and pump waveform, our work offers a promising perspective to achieve a sufficient level of squeezing in a cluster state with a size necessary for MBQC~\cite{walshe_robust_2019}.
While the addition of a non-gaussian resource is needed for full implementation, our work constitutes an important first step toward measurement-based quantum computation on a continuous-variable system at microwave frequencies.

\begin{acknowledgments}
We acknowledge the National Institute of Standards and Technology (NIST) for providing the JPA used in this experiment.
We thank G. Ferrini, J. Aumentado, and J.D. Teufel for fruitful discussions.
This work was supported by the Knut and Alice Wallenberg Foundation through the Wallenberg Center for Quantum Technology (WACQT).
\end{acknowledgments}

\bibliography{Refs}
\nocite{pfister_continuous-variable_2020,tholen_measurement_2022,malnou_three-wave_2021,holevo_evaluating_2001, weedbrook_gaussian_2012,petrovnin_generation_2023, mariantoni_planck_2010, clerk_introduction_2010, simon_quantum-noise_1994, shchukin_recovering_2016, mallet_quantum_2011, jolin_multipartite_2023, ou_realization_1992, braunstein_quantum_2005, yamamoto_principles_2016}

\end{document}


\preprint{APS/123-QED}

\title{Supplemental material: Continuous-variable square-ladder cluster states in a microwave frequency comb} 

\author{Fabio Lingua}
 \email{lingua@kth.se}
 \affiliation{Department of Applied Physics, KTH Royal Institute of Technology, SE-10691 Stockholm, Sweden}
 
\author{J. C. Rivera Hernández}
 \affiliation{Department of Applied Physics, KTH Royal Institute of Technology, SE-10691 Stockholm, Sweden}

\author{Michele Cortinovis}
 \affiliation{Department of Applied Physics, KTH Royal Institute of Technology, SE-10691 Stockholm, Sweden}
 \affiliation{Dipartimento di Fisica, Politecnico di Milano, I-20133 Milano, Italy}

\author{David B. Haviland}
 \affiliation{Department of Applied Physics, KTH Royal Institute of Technology, SE-10691 Stockholm, Sweden}

\date{\today}

\maketitle

\section{Rotation of Covariance Matrix}

The covariance matrix is created from the measured quadrature data
\begin{equation}
    V=\langle\Delta\vec{q} \cdot \Delta\vec{q}^T\rangle,
\end{equation}
where $\Delta\vec{q}= \vec{q} -\expval{\vec{q}}$, $\vec{q}=(x_1,p_1,x_2,p_2,\dots,x_i,p_i,\dots )^T$ is the column vector containing the quadratures of each mode and $\expval{\vec{q}}$ its expectation value. 

Let us define the rotation matrix of a single quadrature pair by an angle $\theta$ as
\begin{equation}
    R_\theta=\begin{pmatrix}
    \cos\theta & -\sin\theta\\
    \sin\theta & \cos\theta
    \end{pmatrix},
\end{equation}
with unitary properties $R_{\theta}^{-1} = R_{-\theta}$, $R_{\theta}^{-1}\cdot R_{\theta}=R_{\theta}\cdot R_{\theta}^{-1}=\mathbb{I}$, and $R_{\theta}^{-1} = R_{\theta}^{T} $.
The rotation of the full covariance matrix by the same angle $\theta$ is achieved through the unitary transformation:
\begin{equation}
    V^\prime = U_\theta \cdot V\cdot U_\theta^{T}
    \label{eq:Vrot}
\end{equation}
where
\begin{equation}
    U_\theta=\mathbb{I}_n\otimes R_{\theta}=\begin{pmatrix}
    R_\theta&&&&\\
    &R_\theta&&&\\
    &&R_\theta&&\\
    &&&\ddots&\\
    &&&&R_\theta
    \end{pmatrix}
\end{equation}

The expression~\eqref{eq:Vrot} indicates that rotating the covariance matrix is equivalent to rotating the quadrature vector $\vec{q}$:
\begin{equation}
    V^\prime = U_\theta \cdot V\cdot U_\theta^{T} = U_\theta \Delta\vec{q} \cdot \Delta\vec{q}^{\,T} U_\theta^{T}= \Delta\vec{q}^{\,\prime} \cdot\Delta\vec{q}^{\, \prime \,T}
\end{equation}
where we used the definition for the rotated quadrature vectors $\Delta\vec{q}^{\,\prime}= U_\theta \cdot \vec{q}$ and $\Delta\vec{q}^{\, \prime \,T}= \left(U_\theta \cdot \Delta\vec{q}\right)^T={\Delta\vec{q}^{\,T} \cdot U_\theta^{T}}$.
Rotating the covariance matrix is thus equivalent to rotating the quadrature basis, representing a change in the measurement reference frame.

\section{Cluster States and Nullifiers}

Continuous-variable (CV) cluster states are multi-partite entangled states defined by a canonical graph between $n$ \emph{qumodes}.
These states are continuous superpositions of $n$-mode position states with a CV ${C_{z, ij}=e^{i\alpha_{ij} \hat{x}_i \hat{x}_j}}$ gate applied between mode pairs corresponding to the edges of the canonical graph:
\begin{multline}
    \ket{\Psi}=Z_{G}\ket{0}_{p_1}\ket{0}_{p_2}\dots\ket{0}_{p_n}=\\
    =\frac{1}{2\pi}\iint d\vec{x} \prod_{i,j\in G} e^{i\alpha_{ij} x_i x_j} \ket{x_1}\ket{x_2}\dots\ket{x_n},
    \label{CS_def}
\end{multline}
where $\ket{0}_{p_1}\ket{0}_{p_2}\dots\ket{0}_{p_n}$ is the vacuum in the momentum basis, $d\vec{x} \equiv dx_1 dx_2\dots dx_n$, and
$Z_{G}=\prod_{i,j\in G} C_{z,\, ij}$ contains the sequence of pairwise $C_z$ gate correlations describing the canonical graph $G$.

\begin{figure}
    \includegraphics[width=\columnwidth]{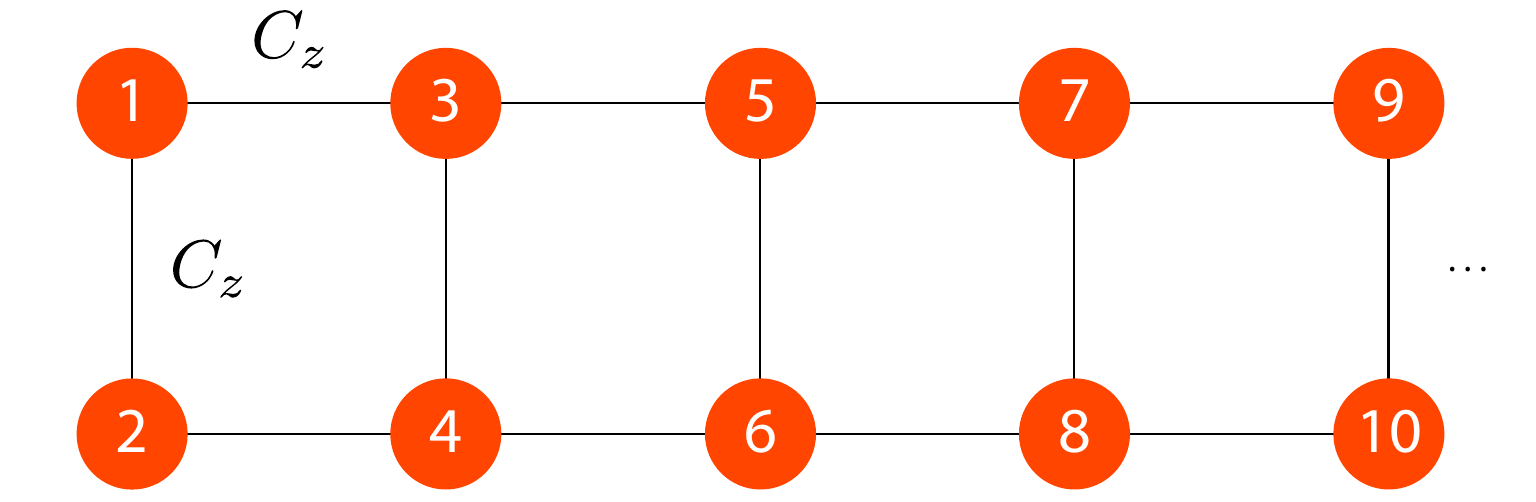}
    \caption{\label{fig:CState} Example of a square-ladder cluster state. Links between nodes correspond to CV $C_z$ gate operations. 
    }
\end{figure}

To realize quantum computation the canonical graph should not be fully connected. 
A graph for measurement-based quantum computation (MBQC) typically features nearest-neighbor connectivity and a lattice topology (see Fig.~\ref{fig:CState} for an example).
Cluster states are defined by their symmetry properties, described by the multiplicative stabilizer group $\mathcal{S}$.
For a stabilizer operator $S\in\mathcal{S}$, the cluster state $\ket{\Psi}$ is invariant under transformation by $S$: 
\begin{equation}
    S\ket{\Psi}=\ket{\Psi} 
    \label{Stab_eq}
\end{equation}
i.e. $\ket{\Psi}$ is an eigenstate of $S$ with eigenvalue $1$. $S$ is in general defined in terms of local stabilizers $S_i$
\begin{equation}
S= \bigoplus_{i=1}^n S_i
\end{equation}
where the direct sum on $i$ runs over all the number of modes.
Local stabilizers $S_i$ obey~\eqref{Stab_eq} and take the form
\begin{equation}
    S_i=X_i\bigotimes_{j\in G_i} Z_j \label{Sdef}
\end{equation}
where $X_i=e^{-i\xi_i \hat{p}_i}$, $Z_j=e^{i\varpi_j \hat{x}_j}$ are the CV analogous of Pauli gates~\cite{pfister_continuous-variable_2020}, $\xi_i,\varpi_j\in\mathbb{R}$ and $G_i$ the graph of modes connected to $i$. 
Note that in Eq.~\eqref{Sdef} the quadrature operators $\hat{p}_i$, $\hat{x}_j$ commute for $i\neq j$. 
One can therefore treat the product of exponential operators as the exponential of the sum.
It follows for Eq.~\eqref{Stab_eq} to be satisfied,
\begin{equation}
    -i\xi_i\left[\hat{p}_i - \sum_{j\in G_i}\frac{\varpi_j}{\xi_i} \hat{x}_j\right]\ket{\Psi}=0 \label{Stab_Null}
\end{equation}
From~\eqref{Stab_Null} we define the \emph{nullifier} for mode $i$ as
\begin{equation}
    N_i=\hat{p}_i - \sum_{j\in G_i}h_{ij} \hat{x}_j
\end{equation}
where $h_{ij}=\frac{\varpi_j}{\xi_i}\in\mathbb{R}$.
A suitable choice of $h_{ij}$ fixes the form of the CV quantum gates and yields the proper stabilization of $\ket{\Psi}$ (see the section below for details).
In the limit of infinite squeezing, the expectation values and variance of the nullifier operators satisfy the relations
\begin{eqnarray}
    &\expval{N_i}&=\bra{\Psi}N_i\ket{\Psi} = 0\\
    &\Delta N_i^2 &= \bra{\Psi}(N_i - \expval{N_i})^2\ket{\Psi} = 0.
    \label{DNi2}
\end{eqnarray}
For finite squeezing it is sufficient to show that $\expval{N_i}\simeq 0$, and $\Delta N_i^2<\Delta N_{i\;0}^2 $, where $\Delta N_{i\;0}^2=\bra{0}(N_i - \expval{N_i})^2\ket{0}$ is the vacuum fluctuations of the nullifier, i.e. the variance of $N_i$ computed for the vacuum, or in the absence of the entangling pumps.    
In our experiment we compute the squeezing of the nullifiers in dB following the standard definition $10\log_{10}( \Delta N_i^2/\Delta N_{i\;0}^2) $.
The nullifier test is considered passed if the ratio $\Delta N_i^2/\Delta N_{i\;0}^2$ is below \SI{0}{\dB} for all modes $i\in G$.

Explicitly, Eq.~\eqref{DNi2} reads
\begin{multline}
    \Delta N_i^2 = \Delta \hat{p}_i^2 + \sum_{jk}\Delta \hat{x}_j\Delta\hat{x}_k +\\ 
    -\sum_{j}\left(\Delta\hat{p}_i\Delta \hat{x}_j + \Delta \hat{x}_j\Delta\hat{p}_i\right) .
    \label{eq:DNi2_xplct}
\end{multline}
Equation~\eqref{eq:DNi2_xplct} is readily evaluated from the elements of the covariance matrix $V$.

\subsection{Cluster State Stabilization Condition}
To prove the stabilization condition of $\ket{\Psi}$, one has to prove that there exists a stabilizer $S_i$ $\forall i\in G$.
Explicitly writing $S_i$ acting on the definition of the cluster state~\eqref{CS_def} and using~\eqref{Sdef} one obtains 
\begin{widetext}
\begin{multline}
    S_i\ket{\Psi}=S_iZ_{G}\ket{0}_{p_1}\dots\ket{0}_{p_i}\dots\ket{0}_{p_n} = \frac{1}{2\pi}\iint d\vec{x}\; e^{-i\xi_i\left[\hat{p}_i - \sum\limits_{j\in G_i}h_{ij} \hat{x}_j\right]}\prod_{k,j\in G} e^{i\alpha_{kj} x_k x_j} \ket{x_1}\dots\ket{x_i}\dots\ket{x_n}=\\
    =\frac{1}{2\pi}\iint d\vec{x}\; e^{-i\xi_i\left[\hat{p}_i - \sum\limits_{j\in G_i}h_{ij} \hat{x}_j\right]}\prod_{k=i,j\in G_i}e^{i\alpha_{ij} x_i x_j}\prod_{k,j\in G-G_i} e^{i\alpha_{kj} x_k x_j} \ket{x_1}\dots\ket{x_i}\dots\ket{x_n}=\\
    =\frac{1}{2\pi}\iint d\vec{x}\; e^{-i\xi_i\hat{p}_i} e^{i\left[\xi_i\sum\limits_{j\in G_i}h_{ij} x_j + \sum\limits_{j\in G_i}\alpha_{ij} x_i x_j\right]}\prod_{k,j\in G-G_i} e^{i\alpha_{kj} x_k x_j} \ket{x_1}\dots\ket{x_i}\dots\ket{x_n}
    \label{Stab_Psi}    
\end{multline}
\end{widetext}
where we exploited the fact that $\ket{x_j}$ is eigenstate of $\hat{x}_j$ with eigenvalue $x_j$, and used the properties $e^{i\beta\hat{x}_j}\ket{x_j}=e^{i\beta x_j}\ket{x_j}$ and $[x_k, x_j]=0$ $\forall j,k$.
The momentum operator in the exponent is a generator of translation $e^{-i\xi_i\hat{p}_i}\ket{x_i}=\ket{x_i + \xi_i}$ from which it follows
\begin{widetext}
\begin{equation}
    \frac{1}{2\pi}\iint d\vec{x}\;  e^{i\left[\xi_i\sum\limits_{j\in G_i}h_{ij} x_j + \sum\limits_{j\in G_i}\alpha_{ij} x_i x_j\right]}\prod_{k,j\in G-G_i} e^{i\alpha_{kj} x_k x_j} \ket{x_1}\dots\ket{x_i + \xi_i}\dots\ket{x_n}.
    \label{Stab_Psi1}    
\end{equation}
\end{widetext}
Changing the variable of integration $x_i\longrightarrow x_i - \xi_i$ results in
\begin{widetext}
\begin{multline}
    \frac{1}{2\pi}\iint d\vec{x}\;  e^{i\sum\limits_{j\in G_i}\left[\xi_i h_{ij} x_j + \alpha_{ij}(x_i - \xi_i) x_j\right]}\prod_{k,j\in G-G_i} e^{i\alpha_{kj} x_k x_j} \ket{x_1}\dots\ket{x_i }\dots\ket{x_n}=\\
    =\frac{1}{2\pi}\iint d\vec{x}\;  e^{i\sum\limits_{j\in G_i}\xi_i\left( h_{ij} - \alpha_{ij}\right)x_j}\cdot e^{i\sum\limits_{j\in G_i} \alpha_{ij}x_i x_j}\prod_{k,j\in G-G_i} e^{i\alpha_{kj} x_k x_j} \ket{x_1}\dots\ket{x_i }\dots\ket{x_n}=\\
    =\frac{1}{2\pi}\iint d\vec{x}\;  e^{i\sum\limits_{j\in G_i}\xi_i\left( h_{ij} - \alpha_{ij}\right)x_j}\cdot \prod_{k,j\in G} e^{i\alpha_{kj} x_k x_j} \ket{x_1}\dots\ket{x_i }\dots\ket{x_n}
    \label{Stab_Psi2}    
\end{multline}
\end{widetext}
For Eq.~\eqref{Stab_Psi2} to be equal to Eq.~\eqref{CS_def} one requires
\begin{equation}
     e^{i\sum\limits_{j\in G_i}\xi_i\left( h_{ij} - \alpha_{ij}\right)x_j}=1,
     \label{cond_large}
\end{equation}
which is certainly satisfied for
\begin{equation}
    h_{ij} = \alpha_{ij}.
    \label{cond_stringent}
\end{equation}
Conditions~\eqref{cond_large}~and~\eqref{cond_stringent} essentially prove the symmetry transformation $S_i\ket{\Psi}=\ket{\Psi}$.
The particular choice of $h_{ij}$  satisfying~\eqref{cond_stringent} fixes the definition of gates $X$, $Z$ in the MBQC toolbox.

\section{Experimental setup}
\begin{figure}[t]
    \includegraphics[width=\columnwidth]{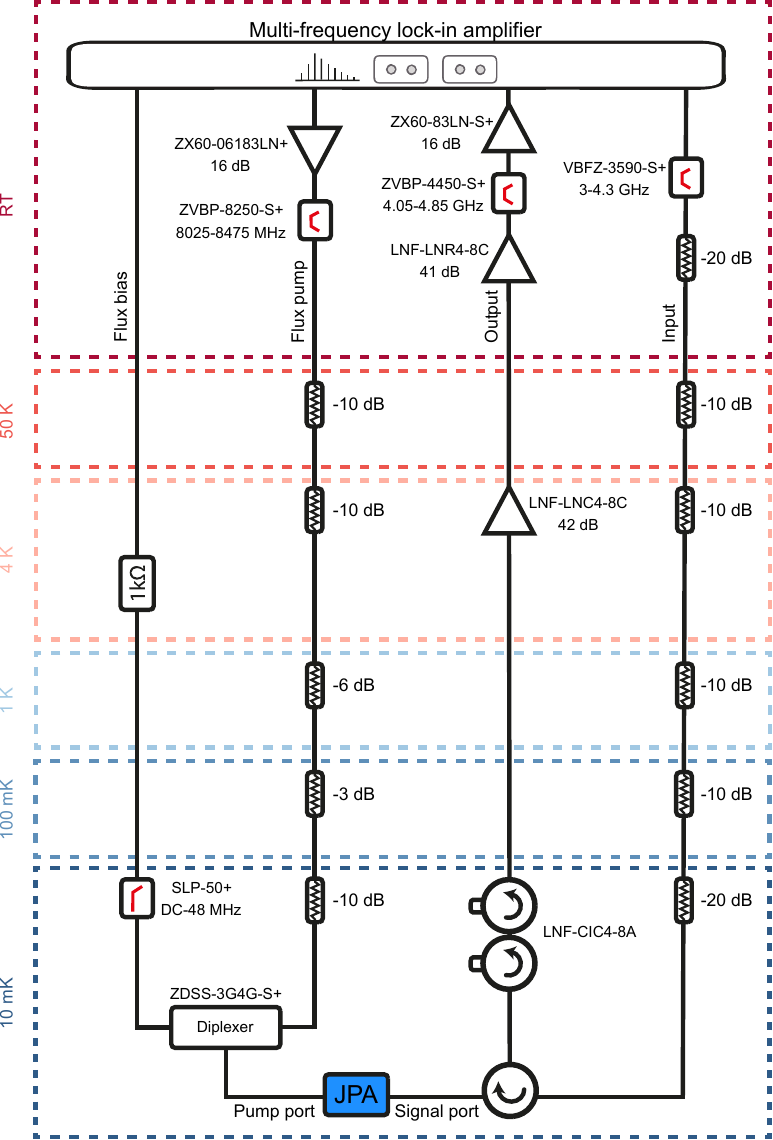}
    \caption{Schematic of the measurement setup, including room temperature and cryogenic electronics.}
    \label{fig:setup}
\end{figure}

Figure~\ref{fig:setup} presents a schematic of the measurement setup. 
The JPA is cooled to \SI{10}{\milli\kelvin} in a Bluefors LD250 dry dilution refrigerator.  
Appropriate attenuation of the high frequency transmission lines at multiple temperature stages ensures that the pump and drive signals are coherent states.
The added noise from the amplifier at \SI{4}{\kelvin} determines the measurement signal-to-noise ratio. 
A double isolator attenuates the amplifier noise back-acting on the sample, ensuring that vacuum fluctuations are present at the signal port.
The JPA operates in reflection, utilizing a circulator to separate the input and output fields.

The multi-frequency lock-in amplifier is a particular firmware running on a microwave digital platform created from a radio-frequency system-on-a-chip~\cite{tholen_measurement_2022}.
The instrument provides the probe signal (for measuring the scattering matrix) and the multi-frequency pump waveform through separate output ports.
All signals across all output ports and all input ports are synchronized through one master clock. 
This instrument also supplies the DC bias for the JPA, which is combined with the pump via a diplexer at base temperature. 
The probe signal in the \qtyrange[range-units=single]{3}{6}{\giga\hertz} band, and the multi-frequency pump waveform in the \qtyrange[range-units=single]{5}{10}{\giga\hertz} band are created with direct digital synthesis utilizing high-speed digital-to-analog converters operating in the second Nyquist zone. 
Demodulated signals in a higher Nyquist zone of the analog-to-digital converter are under-sampled, appearing at aliased frequencies in the 1st Nyquist zone. 
Analog filters are used at room temperature to select the appropriate Nyquist zone. 
This direct digital approach eliminates the gain and phase errors associated with analog IQ mixers.
An essential feature of this instrument is the synchronization of output and input ports and the frequency tuning which establishes the orthogonal mode basis and eliminates Fourier leakage between modes.
\begin{figure*}[t]
    \centering
    \includegraphics[width=\linewidth]{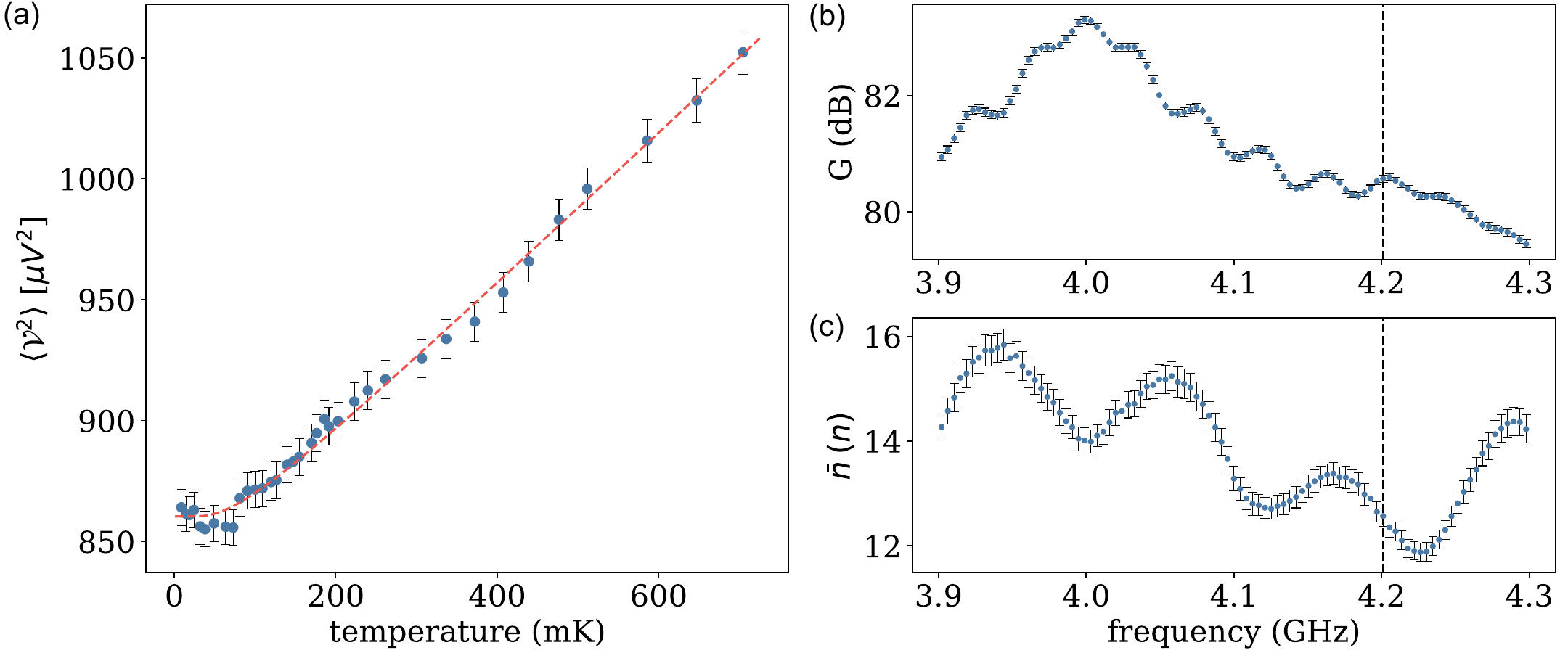}
    \caption{(a) Variance of the voltage noise at \SI{4.2}{\giga\hertz} measured as a function of temperature. The red dashed line corresponds to the fit to Eq.~\eqref{eq:planck-v}. Such a fit at each frequency provides the gain $G$ and the added number of noise photons $\bar{n}$, plotted as a function of frequency in (b) and (c) respectively. The black error bars in all figures refer to the measurement uncertainty.}
    \label{fig:planck}
\end{figure*}

\section{Covariance Matrix Reconstruction}
To analyze the quantum correlations induced by the JPA we must remove the uncorrelated classical noise added by our amplification chain.
This chain acts as a noisy bosonic Gaussian channel~\cite{malnou_three-wave_2021}, where each mode $i$ experiences a frequency-dependent gain $G(\omega_i)$ and an average number of added noise photons $\bar{n}_i$.
The effect of a bosonic Gaussian channel on an arbitrary Gaussian state transforms the mean and covariance matrix as follows~\cite{holevo_evaluating_2001, weedbrook_gaussian_2012}
\begin{equation}
    \bar{x} \rightarrow T\bar{x} + d, \;\;\;\;\; V \rightarrow TVT^T + N,
    \label{eq:Gaussian-channel}
\end{equation}
where $T=\bigoplus_i \sqrt{G(\omega_i)}I_i$ and $N = \bigoplus_i \left(G(\omega_i) - 1\right)(\bar{n}_i + 1/2)I_i$.
Since $T$ and $N$ are diagonal matrices, the resulting measured covariance matrix $V^\text{meas}$ is
\begin{equation}
    V^\text{meas} = T^2 V^\text{quant} + N,
    \label{eq:V-trans}
\end{equation}
where $V^\text{quant}$ represents the covariance matrix of the quantum state scattered by the JPA before amplification.

Unavoidable uncertainties in estimating the added noise $N$ result in unphysical covariance matrices due to excessive noise subtraction.
To circumvent this problem we exploit the fact (verified by experiment) that the amplification chain's added classical noise at different frequencies is completely uncorrelated.
This added noise therefore only contributes to the diagonal elements of $V^\text{meas}$.
In the absence of pumps, the ratio between the quantum noise and the added classical noise on the diagonal is
\begin{equation}
    \dfrac{V^\text{quant}_{0\,ii}}{V^\text{classic}_{0\,ii}} = \dfrac{1}{2\bar{n}_i + 1}.
    \label{eq:ratio-Vqc-off}
\end{equation}
The measured covariance matrix with the pump off $V^\text{meas}_0$ after amplification is given by
\begin{equation}
    V^\text{meas}_{0\,ii} = G(\omega_i) \left(V^\text{quant}_{0\,ii} + V^\text{classic}_{0\,ii}\right).
    \label{eq:Vm0-diag}
\end{equation}
Using Eqs.~\eqref{eq:ratio-Vqc-off}~and~\eqref{eq:Vm0-diag}, and assuming that the classical uncorrelated noise remains the same for the pump-on case, we recover the quantum covariance matrix as~\cite{petrovnin_generation_2023}: 
\begin{equation}
    V^\text{quant} = T^{-2} \left( V^\text{meas} - V^\text{meas}_0 \right) + V^\text{quant}_0, 
    \label{eq:Vq-rec}
\end{equation}
where $V^\text{quant}_0$ can be calculated from $V^\text{meas}_0$ as
\begin{align}
    V^\text{quant}_{0\,ij} = 
    \begin{cases} 
        \dfrac{1}{2\bar{n}_i+2} \; G(\omega_i)^{-1} \; V^\text{meas}_{0\,ij}, \;\; \text{for} \;\; i=j \\ \\
        G(\omega_i)^{-1} \; V^\text{meas}_{0\,ij}, \;\; \text{for} \;\; i \neq j
    \end{cases}
    \label{eq:Vqoff}
\end{align}

We estimate $G(\omega_i)$ and $\bar{n}_i$ by measuring the Johnson-Nyquist noise spectral density emitted by a \SI{50}{\ohm} source (the isolator) as a function of temperature $T_\mathrm{mxc}$.
This method, known as Planck spectroscopy~\cite{mariantoni_planck_2010}, involves slowly heating the mixing chamber to vary the isolator's temperature and allowing \SI{2}{\hour} for thermalization before measuring.
The amplified power spectral density of the noise is given by~\cite{clerk_introduction_2010}
\begin{equation}
    P = \dfrac{1}{2} G(\omega_i) hf_i \left[ \coth{\left( \dfrac{hf_i}{2k_B T_\mathrm{mxc}} \right)} + 2\bar{n}_i + 1 \right].
    \label{eq:planck}
\end{equation}
In terms of the voltage variance over a measurement bandwidth $\Delta$, we have
\begin{equation}
    \langle \mathcal{V}^2 \rangle = 4 \Delta P \times \SI{50}{\ohm}.
    \label{eq:planck-v}
\end{equation}
Figure~\ref{fig:planck} shows the fit of Eq.~\eqref{eq:planck-v} to the measured voltage variance. 
We measure at 192 frequencies simultaneously using the multifrequency lock-in~\cite{tholen_measurement_2022}. 
Performing this fit at each frequency, we extract the gain $G(\omega_i)$ and added noise $\bar{n}_i$ as a function of frequency, as shown in Fig.~\ref{fig:planck}b and Fig.~\ref{fig:planck}c, respectively.

Inherent difficulties associated with extracting the quantum noise from the added noise, as well as uncertainties in our calibration, lead to a reconstructed $V^\text{quant}$ which is nonphysical, in the sense that it violates the Heisenberg uncertainty principle. 
The uncertainty principle for a covariance matrix expressed in photon number ($\Delta x \Delta p \geq 1/2$ or $\hbar=1$) is stated as~\cite{simon_quantum-noise_1994, weedbrook_gaussian_2012}
\begin{equation}
    V^\text{quant} \pm \dfrac{i}{2} \Omega \geq 0, \;\;\;\; \text{with} \;\; \Omega = \bigoplus_i \begin{pmatrix}
                    0 & 1\\
                    -1 & 0
                \end{pmatrix}.
    \label{eq:physicality}
\end{equation}

To compute the best-approximated physical covariance matrix $V$, we perform a constrained minimization of an objective function~\cite{shchukin_recovering_2016}
\begin{equation}
    \min_{V} \left( \max_{ij}\frac{ \left|V^\text{quant}_{ij} - V_{ij} \right|}{\sigma^{V^\text{quant}}_{ij}} \right),
    \label{eq:shchukin}
\end{equation}
where $\sigma^{V^\text{quant}}_{ij}$ represents the experimental error associated with the $ij$-th element of $V^\text{quant}$.

\section{Error Analysis}
To compute the best approximated physical covariance matrix and estimate the uncertainty in the nullifiers, we need the experimental errors in the quantum covariance matrix $\sigma^{V^\text{quant}}_{ij}$.
We estimate these errors by combining the uncertainties in the calibration and measurement through error propagation of Eq.~\eqref{eq:Vq-rec}.
The uncertainty of each element of $V^\text{quant}$ is given by:
\begin{widetext}
\begin{multline}
    \sigma^{V^\text{quant}}_{ij} = \sqrt{\left( \dfrac{\partial V^\text{quant}_{ij}}{\partial G(\omega_i)} \right)^2 \sigma_{G(\omega_i)}^2 + \left( \dfrac{\partial V^\text{quant}_{ij}}{\partial V^\text{meas}_{ij}} \right)^2 \sigma_{V^\text{meas}_{ij}}^2 + \left( \dfrac{\partial V^\text{quant}_{ij}}{\partial V^\text{meas}_{0\,ij}} \right)^2 \sigma_{V^\text{meas}_{0\,ij}}^2 + \left( \dfrac{\partial V^\text{quant}_{ij}}{\partial V^\text{quant}_{0\,ij}} \right)^2 \sigma_{V^\text{quant}_{0\,ij}}^2} = \\
    = \sqrt{\left| \dfrac{(V^\text{meas}_{ij} - V^\text{meas}_{0\,ij})}{G(\omega_i)^{2}} \right|^2 \sigma_{G(\omega_i)}^2 + G(\omega_i)^{-2} \left( \sigma_{V^\text{meas}_{ij}}^2 + \sigma_{V^\text{meas}_{0\,ij}}^2 \right) + \sigma_{V^\text{quant}_{0\,ij}}^2},
    \label{eq:error-Vq}
\end{multline}
\end{widetext}
where $\sigma_{G(\omega_i)}$ is the uncertainty in $G(\omega_i)$, $\sigma_{V^\text{meas}}$ and $\sigma_{V^\text{meas}_0}$ are the measurement errors in the measured covariance matrices with and without pumps, respectively, and $\sigma_{V^\text{quant}_0}$ are the errors in the pump-off quantum covariance matrix.
To estimate $\sigma_{V^\text{quant}_0}$ we apply the chain rule to Eqs.~\eqref{eq:Vq-rec}~and~\eqref{eq:Vqoff}, obtaining 
\begin{widetext}
\begin{equation}
    \sigma^{V^\text{quant}}_{0\,ij} = \sqrt{\left( \dfrac{\partial V^\text{quant}_{0\,ij}}{\partial G(\omega_i)} \right)^2 \sigma_{G(\omega_i)}^2 + \left( \dfrac{\partial V^\text{quant}_{0\,ij}}{\partial \bar{n}_i} \right)^2 \sigma_{\bar{n}_i}^2 + \left( \dfrac{\partial V^\text{quant}_{0\,ij}}{\partial V^\text{meas}_{0\,ij}} \right)^2 \sigma_{V^\text{meas}_{0\,ij}}^2 + 2 \left( \dfrac{\partial V^\text{quant}_{0\,ij}}{\partial G(\omega_i)} \right) \left( \dfrac{\partial V^\text{quant}_{0\,ij}}{\partial \bar{n}_i} \right) \sigma_{G(\omega_i), \bar{n}_i}}
\end{equation}
that for $i=j$ it reads:
\begin{equation}
    \sigma^{V^\text{quant}}_{0\,ii} = \sqrt{ \left(\dfrac{G(\omega_i)^{-2} V^\text{meas}_{0\,ij}}{2\bar{n}_i + 2}\right)^2 \sigma_{G(\omega_i)}^2 + \left(\dfrac{G(\omega_i)^{-1} V^\text{meas}_{0\,ij}}{2(\bar{n}_i + 1)^2} \right)^2 \sigma_{\bar{n}_i}^2 + \left( \dfrac{G(\omega_i)^{-1}}{2\bar{n}_i + 2} \right)^2 \sigma_{V^\text{meas}_{0\,ij}}^2 - \dfrac{G(\omega_i)^{-3} \left(V^\text{meas}_{0\,ij}\right)^2}{2(\bar{n}_i + 1)^3} \sigma_{G(\omega_i), \bar{n}_i} },  \nonumber
\end{equation}
while for $i \ne j$ it becomes:
\begin{equation}
    \sigma^{V^\text{quant}}_{0\,ij} = \sqrt{ \left(G(\omega_i)^{-2} V^\text{meas}_{0\,ij}\right)^2 \sigma_{G(\omega_i)}^2 + G(\omega_i)^{-2} \sigma_{V^\text{meas}_{0\,ij}}^2}.  \nonumber
\end{equation}
For the nullifier, the uncertainty is estimated using error propagation through Eq.~\eqref{eq:DNi2_xplct}:
\begin{equation}
    \sigma_{\Delta N_i}^2 = \sigma_{V^\text{quant}_{i+1\,i+1}}^2 + \sum_{j,k} \sigma_{V^\text{quant}_{jk}}^2 + \sum_{j} \left( \sigma_{V^\text{quant}_{i+1 \, j}}^2 + \sigma_{V^\text{quant}_{j \, i+1}}^2 \right).
    \label{eq:error-nullifier}
\end{equation}
\end{widetext}

\section{One parametric pump}
\begin{figure}
\includegraphics[width=\columnwidth]{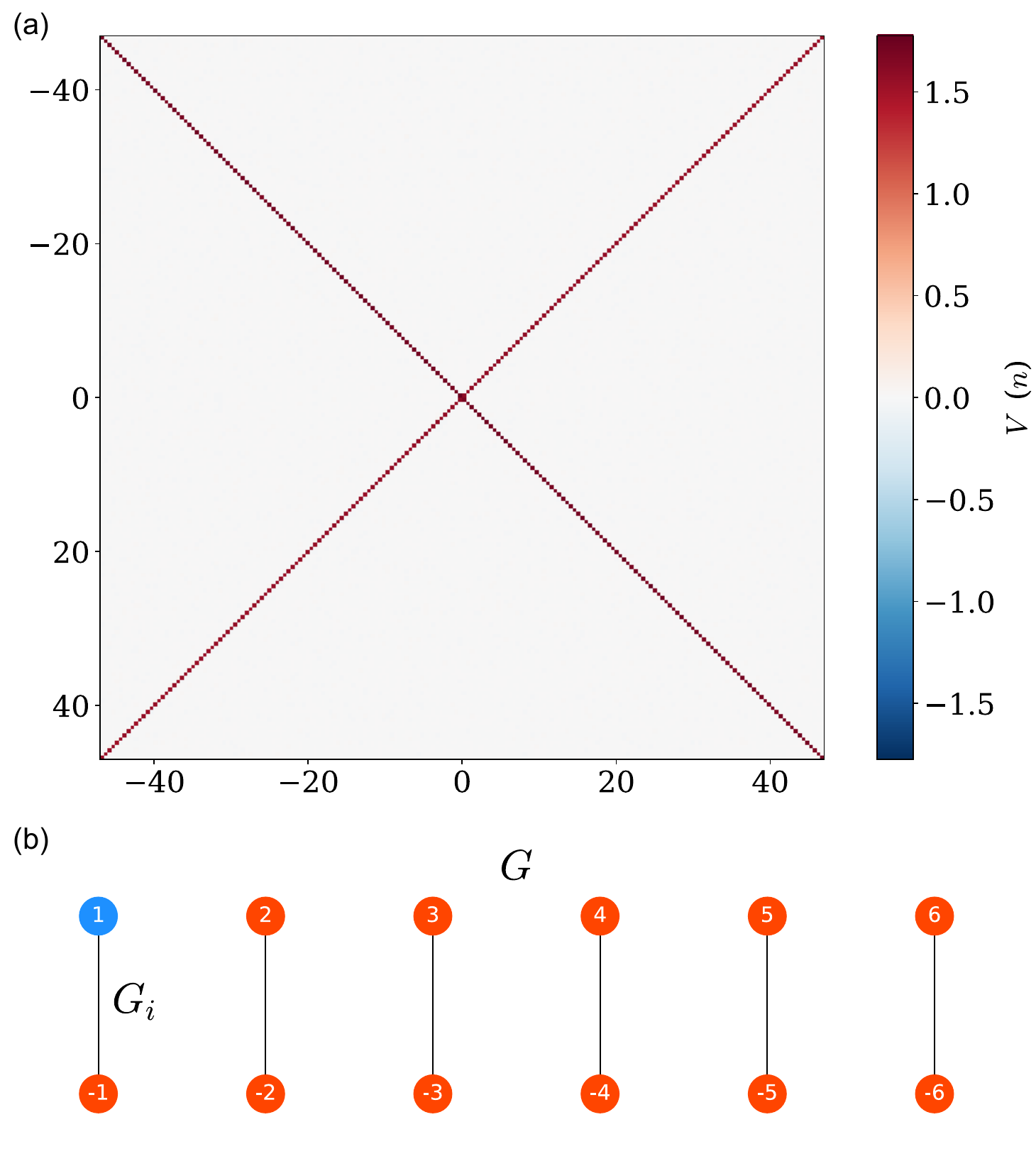}
    \caption{\label{fig:cov1pmp} (a) Covariance matrix $V$ of $n=95$ modes for case of one pump at $g=0.86g_\text{3dB}$ and rotation angle of maximum squeezing $\theta=0.90\pi$. (b) Canonical graph $G$ and single mode graphs $G_i$ representing the two-mode squeezing correlations described by $V$.
    }
\end{figure}
\begin{figure}
\includegraphics[width=\columnwidth]{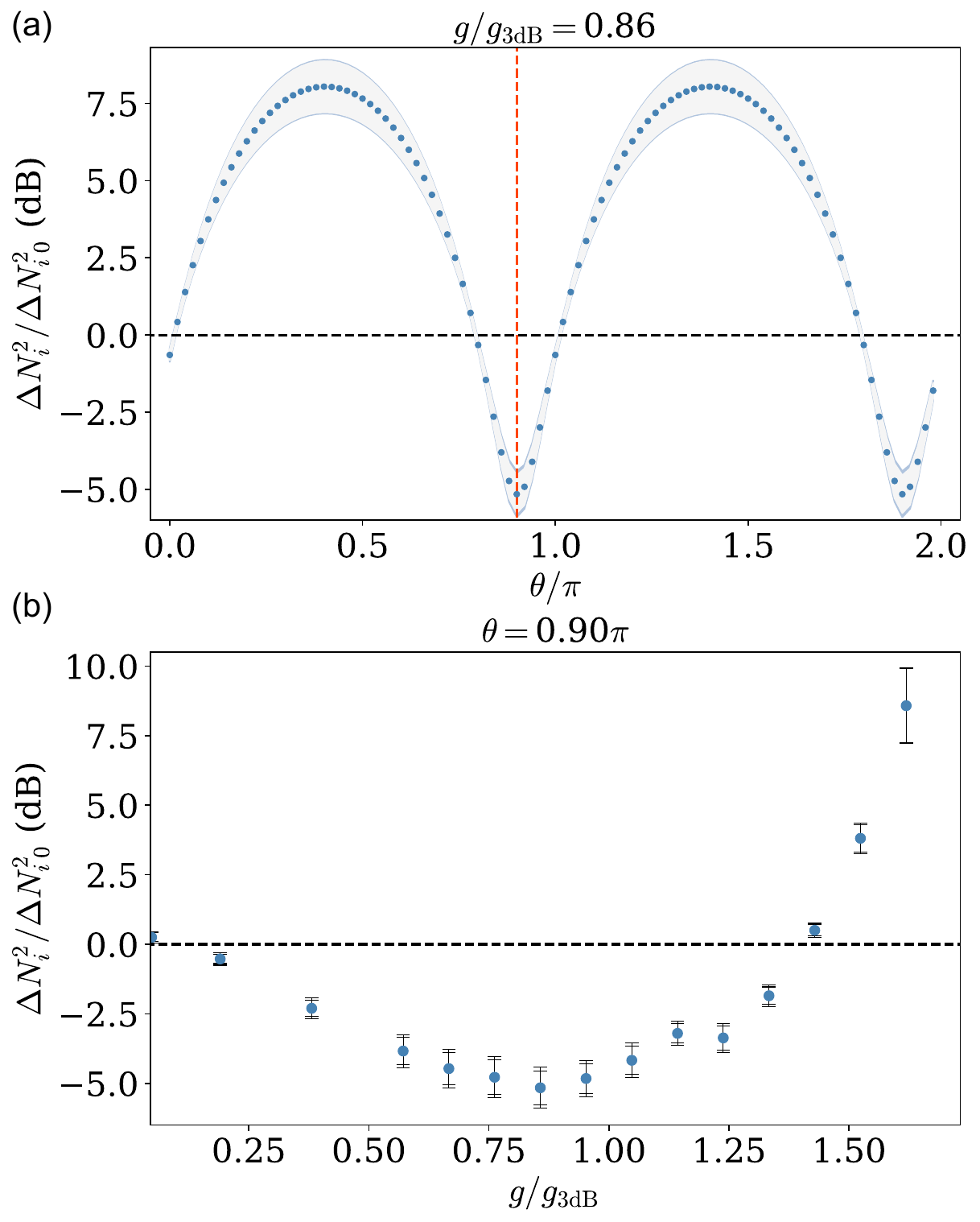}
    \caption{\label{fig:null1pmp} (a) Nullifier variance for the case of one pump normalized to its vacuum value in \SI{}{\dB} and plotted as a function of the global rotation angle $\theta$ for $g=0.86g_\text{3dB}$. (b) Nullifier variance as a function of pump power for $\theta=0.90\pi$, at maximum squeezing [vertical red dashed line in panel (a)]}.
\end{figure}

Here we analyze a variant of the experiment described in the main paper for the specific case of one pump at frequency $2\omega_0$ applied to the JPA. 
In FIG.~\ref{fig:cov1pmp}(a) we show the physical covariance matrix for the one-pump case.
The matrix exhibits the usual diagonal corresponding to $\Delta x_i^2$, $\Delta p_i^2$ of each mode, and an anti-diagonal, corresponding to mode correlations resulting from intermodulation products connecting modes $i$ and $-i$.
The first six single-mode graphs $G_i$ are shown in FIG.~\ref{fig:cov1pmp}(b). 
The canonical correlation graph $G$ of this correlation scheme consists of $(n-1)/2$ pairs of two-mode squeezing correlations~\cite{mallet_quantum_2011, jolin_multipartite_2023}.
The nullifier operator on the single mode graph $G_i$ reduces to
\begin{equation}
    N_i=\hat{p}_i - \hat{x}_{-i},
\label{null1pmp}
\end{equation}
with variance
\begin{equation}
\Delta N_i^2=\Delta\hat{p}_i^2 + \Delta\hat{x}_{-i}^2 - (\Delta\hat{p}_i\Delta\hat{x}_{-i} + \Delta\hat{x}_{-i}\Delta\hat{p}_i).
\label{DNi2_1pmp}
\end{equation}
We verified that $\langle N_i\rangle\simeq 0$ $\forall i\in G$. 
By studying the nullifier variance~\eqref{DNi2_1pmp} as a function of the total rotation angle $\theta$, we find the correct reference frame that maximizes the nullifier squeezing below its vacuum value for $\theta\approx 0.90(k+1)\pi$, $k\in\mathbb{Z}$. 
In FIG.~\ref{fig:null1pmp}(a) we plot $\Delta N_i^2/\Delta N_{i\;0}^2$ averaged over all modes in $G$ in \SI{}{\dB}.
We see that we can reach up to \SI{5}{\dB} of squeezing for a pump power of $g/g_\text{3dB}\simeq0.86$.
FIG.~\ref{fig:null1pmp}(b) for the $\theta$ at maximum squeezing [$\theta$ corresponding to the vertical red dashed line in FIG.~\ref{fig:null1pmp}(b)].
We see that the nullifier is squeezed from $g\geq0$ up to an optimal pumping strength of $g\simeq0.86g_\text{3dB}$ and then starts progressively decreasing for higher pumping powers until $g\simeq1.5g_\text{3dB}$ where it crosses above the vacuum level. 
The existence of a pump power of optimal squeezing is consistent with numerical analysis of the JPA Hamiltonian expanded in the mode basis described in the section~\ref{sec:NumAnalysis} below.
Both the optimal level of squeezing, and its progressive destruction as a function of pump power can be easily explained in terms of losses in the system.

\begin{figure*}[t]
\includegraphics[width=\textwidth]{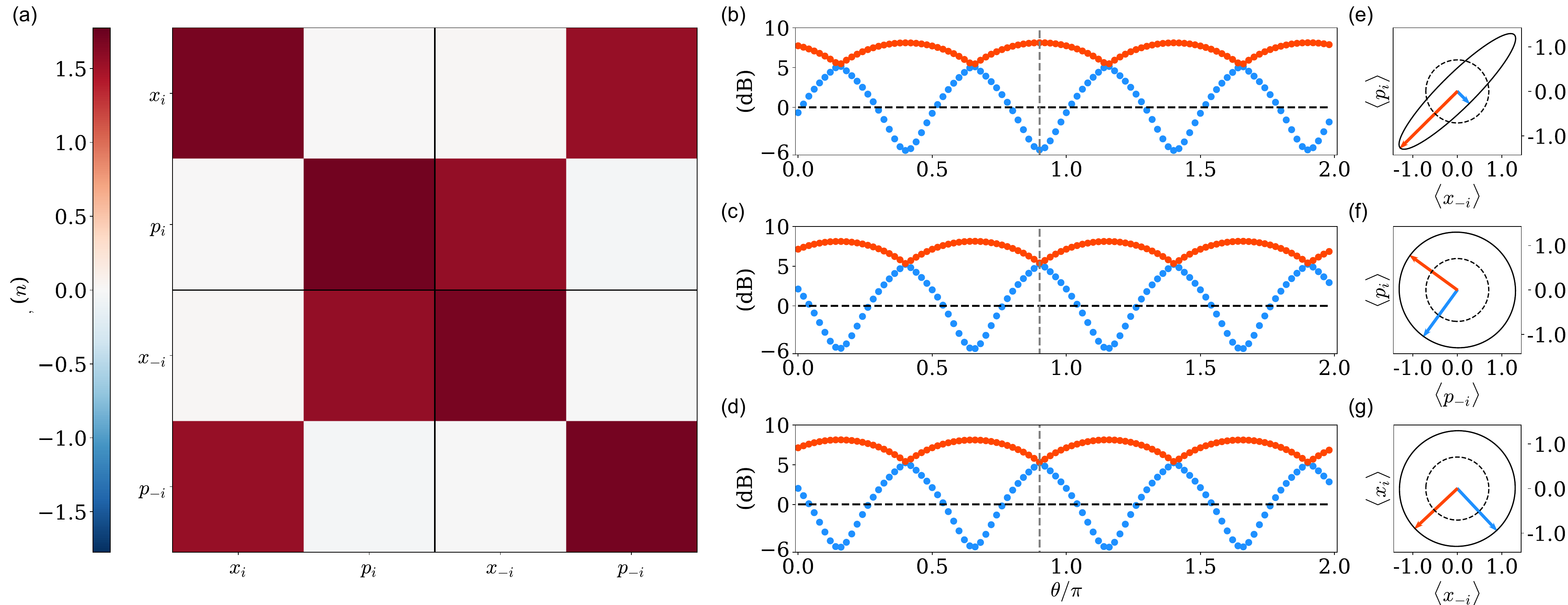}
    \caption{\label{fig:sqz1pmp} (a) Two-mode covariance matrix $V_{i,-i}$ for $\theta=0.90\pi$ [vertical gray dashed line in (b)-(d)]}. Larger (red), smaller (blue) eigenvalues as a function of rotation $\theta$ corresponding to quadratures pairs: $\hat{p}_i$-$\hat{x}_{-i}$ (b), $\hat{p}_i$-$\hat{p}_{-i}$ (c), and $\hat{x}_i$-$\hat{x}_{-i}$ (d). (e)-(g) corresponding squeezing ellipses for $\theta$ corresponding to the vertical gray dashed line in (b)-(d). All panels refer to the pump power $g=0.86g_\text{3dB}$.
\end{figure*}

This measurement is a benchmark for characterizing the maximum level of squeezing with our experimental setup.
In FIG.~\ref{fig:sqz1pmp} we investigate the two-mode squeezing as a function of the global rotation angle $\theta$.
We report results for the generic mode pair $i$, $-i$ as uniformity is observed for the full frequency comb.
Figure~\ref{fig:sqz1pmp}(a) shows the two-mode covariance matrix at $\theta$ of maximum squeezing corresponding to the vertical dashed gray line in panels (b)-(d).
Figure~\ref{fig:sqz1pmp}(b)-(d) shows the larger (red) and smaller (blue) eigenvalues of the three $2 \cross 2$ covariance matrices containing all quadrature correlations: $\hat{p}_i$-$\hat{x}_{-i}$ (b), $\hat{p}_i$-$\hat{p}_{-i}$ (c), and $\hat{x}_i$-$\hat{x}_{-i}$ (d), plotted as a function of the global rotation angle $\theta$ for quadratures.
Panels (e)-(g) show the eigenvectors and the corresponding squeezing ellipses at the value of $\theta$ given by the vertical dashed gray line of panels (b)-(d).
Maximum squeezing in $px$ corresponds to zero squeezing in the $pp$, $xx$ directions, and vice-versa.
We also see that our setup exhibits up to \SI{5.42}{\dB} of squeezing and \SI{10}{dB} of anti-squeezing.
This asymmetry between measured squeezing and anti-squeezing is indicative of some undesirable loss mechanism that is degrading our ability to generate stronger quantum correlations in the microwave frequency comb. 

\subsection{Two-mode squeezing, entanglement, and cluster states}

Two-mode squeezing is the simplest example of CV entanglement, and it constitutes the definition of the seminal EPR state~\cite{ou_realization_1992, braunstein_quantum_2005}, which is a two-mode cluster state.
From the general definition of the cluster state~\eqref{CS_def}, non-factorizability (i.e. the entanglement) follows naturally.
The integration cannot be separated in the product of multiple integrals over each variable, due to the presence of the operator $e^{i\alpha_{ij} \hat{x}_i\hat{x}_j}$.
This operator acts as a ``controlled'' displacement in the momentum space from $\ket{p_i} \longrightarrow \ket{p_i +\alpha_{ij} x_j}$ and physically describes a correlation between the measure of quadratures $p_i$ and $x_j$.
The correlation can be quantified by statistical evaluation of the product $p_i x_j$ between the two quadratures.
Quantum mechanically, this corresponds to the evaluation of the expectation value of the product of the two quadrature operators $\langle \hat{p}_i\hat{x}_j\rangle$.
The computation of this expectation value corresponds to the computation of the covariance matrix element $\Delta \hat p_i \Delta \hat x_j\approx \langle \hat{p}_i\hat{x}_j\rangle$ that, by definition, measures two-mode squeezing between $i$ and $j$.

\subsection{Losses and Squeezing} \label{sec:NumAnalysis}

We perform a numerical analysis of two-mode squeezing as a function of pump power and losses.
The JPA Hamiltonian is expanded in the mode basis and its time evolution is computed by numerical solution of the Lindblad master equation. 
The JPA Hamiltonian is defined as~\cite{yamamoto_principles_2016}
\begin{equation}
    H = \omega_0 \left(A^\dagger A + \frac{1}{2}\right) + \frac{\omega_0}{2} g_p(t)(A^\dagger + A)^2.
\end{equation}
We define the pump signal as characterized as a sum of pure frequency tones
\begin{equation}
    g_p(t) = \sum_k A_k \cos(\Omega_k t + \phi_k) = \sum_k g_k\, e^{i\Omega_k t} + g_k^*\, e^{-i\Omega_k t},    
\end{equation}
where $g_k = \frac{A_k}{2} e^{-i\phi_k}$ is the pump strength.
We focus on a single pump tone at frequency \(\Omega_0 = 2\omega_0\) and complex amplitude $g$.
Expanding the ladder operators on the frequency-mode basis
\begin{equation}
A(t) = \sum_i a_i\, e^{i\omega_it}, \quad A^\dagger(t) = \sum_i a_i^\dagger\, e^{-i\omega_it},    
\end{equation}
and performing a change of reference in the rotating frame of the resonator, we obtain
\begin{align}
    H &= \omega_0 \sum_{i,j} a_i^\dagger a_j \, e^{i(\omega_j - \omega_i)t} +
     \nonumber\\
    &+ \frac{\omega_0}{2}  \sum_{i,j} \left(ge^{i \Omega_0 t} + g^*e^{-i \Omega_0 t}  \right) \left(a_i^\dagger a_j^\dagger + \, a_i a_j \right).
\end{align}
Here we neglect high-frequency terms and assume $\omega_0(1 + g_p(t))\approx\omega_0$ in the weak pump limit.
We consider only two modes equally spaced around the JPA resonance frequency $\omega_0$ with frequency separation spanning the entire bandwidth of the JPA.
The Lindblad master equation describes the dynamics
\begin{equation}
    \dot{\rho} = -\frac{i}{\hbar}[H,\rho] + \sum_{j} \gamma\, \mathcal{D}[a_j]\rho,
\end{equation}
with the dissipative superoperator \(\mathcal{D}[a_j]\rho\) defined as
\begin{equation}
    \mathcal{D}[a_j]\rho = a_j\, \rho\, a_j^\dagger - \frac{1}{2}\{a_j^\dagger a_j,\, \rho\}.
\end{equation}
Here \(\gamma\) is the total loss rate assumed identical for all modes \(j = -1,1\).

\begin{figure}[t]
    \centering
    \includegraphics[width=1\linewidth]{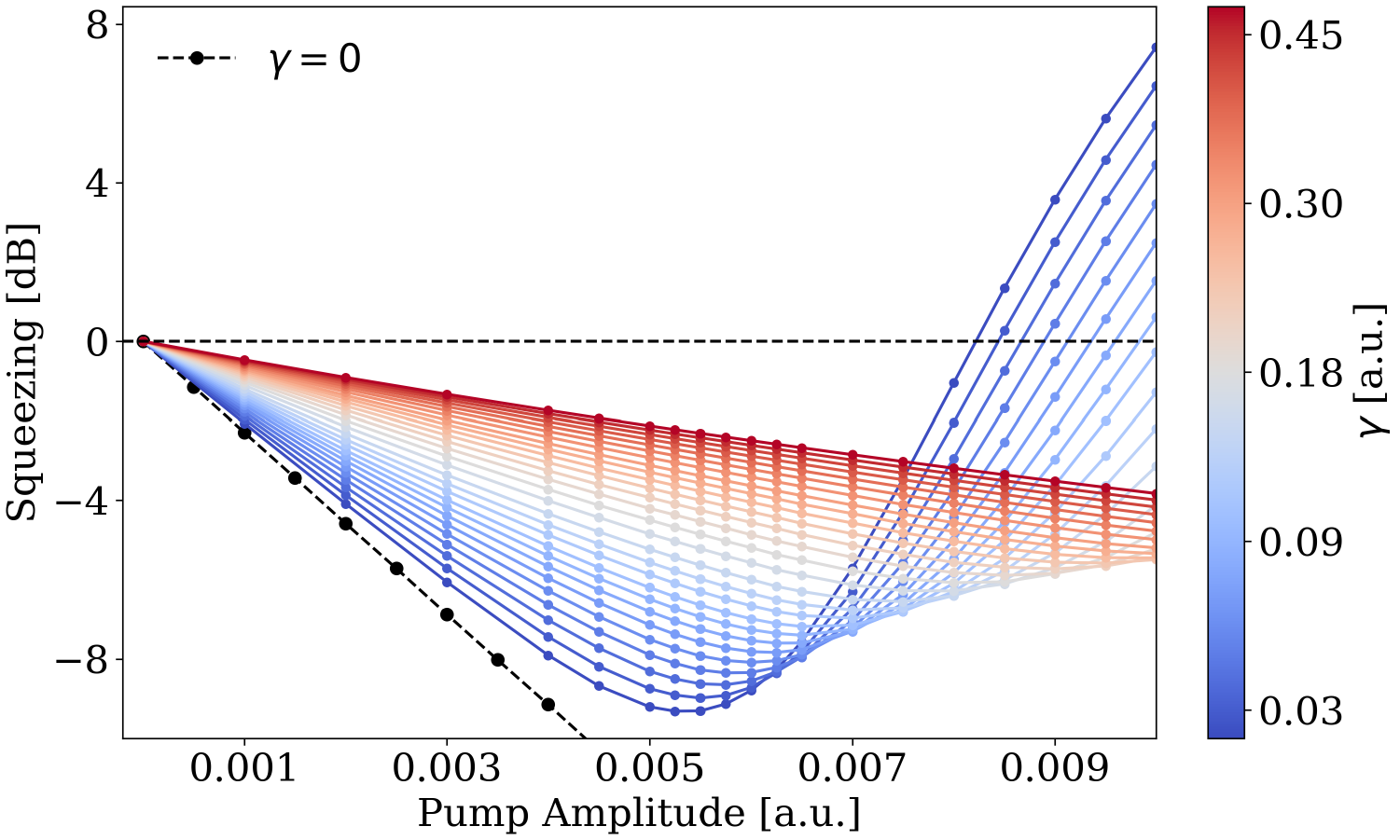}
    \caption{Squeezing as a function of the pump amplitude (in arbitrary units) for different values of the loss rate $\gamma$ indicated by the color scale. The horizontal dashed line marks the vacuum level. The black dashed-dot line refers to the lossless case $\gamma=0$.}
    \label{fig:squeezing_vs_pump_vs_gamma}
\end{figure}

This formulation consistently incorporates both the unitary and dissipative aspects of the system.
Figure~\ref{fig:squeezing_vs_pump_vs_gamma} shows the numerical computation of the squeezing as a function of pump amplitude for various values of \(\gamma\). 
The smallest eigenvalue of the covariance matrix between the quadratures pair \(\hat{p}_i\)-\(\hat{x}_{-i}\) quantifies the squeezing.
The results are evaluated at the $\theta$ of maximum squeezing and normalized to the pump-off case, as in the experiment.
We observe qualitative agreement between the numerical results and the measured two-mode squeezing in FIG.~\ref{fig:null1pmp}.
The presence of losses explains the existence of the minimum at a particular pump amplitude, corresponding to the maximum achievable squeezing.
In contrast, the lossless case (black dashed line in FIG.~\ref{fig:squeezing_vs_pump_vs_gamma}) exhibits a monotonically increased squeezing with pump power.
Furthermore, the numerical solutions for  $\gamma\neq 0$ show degradation of squeezing above the optimal pump amplitude, as observed in the experiment.
Our numerical solutions also confirm an imbalance between the levels of squeezing and anti-squeezing induced by the losses, as seen in Fig.~\ref{fig:sqz1pmp}(b)-(d).

\bibliography{Refs}